\documentclass[11pt]{article}
\usepackage{epsfig}
\usepackage{amssymb}
\newcommand{\be}{\begin{eqnarray}}
\newcommand{\ee}{\end{eqnarray}}
\newcommand{\boldpi}{\mbox{\boldmath $\pi$}}

\newcommand{\boldT}{\mbox{\boldmath $T$}}

\begin{document}
\title{\vspace{-1.cm}
\hfill {\small FZJ--IKP(TH)--2005--22, HISKP-TH-07-19}\\
\vspace{0.3cm}
Role of the $\Delta(1232)$ in pion-deuteron scattering at threshold\\
within chiral effective field theory }
\author{V.~Baru$^a$, J.~Haidenbauer$^b$, C.~Hanhart$^b$, A.~Kudryavtsev$^a$,
\\ V.~Lensky$^b$, and Ulf-G.~Mei{\ss}ner$^{b, c}$
\vspace{0.5cm}\\
{\small $^a$ Institute of Theoretical and Experimental Physics,} \\
{\small 117259, B. Cheremushkinskaya 25, Moscow, Russia} \\
{\small $^b$ Institut f\"{u}r Kernphysik, Forschungszentrum J\"{u}lich GmbH,}\\ 
{\small D--52425 J\"{u}lich, Germany} \\
{\small $^c$ Helmholtz-Institut f\"{u}r Strahlen- und Kernphysik (Theorie), } \\ 
{\small Universit\"at Bonn, Nu{\ss}allee 14-16, D--53115 Bonn, Germany }
}
\maketitle
\begin{abstract}
We investigate the role of the delta isobar in the reaction
$\pi d\to \pi d$ at threshold in chiral effective field theory. 
We discuss the corresponding power counting and argue
that this calculation  completes the evaluation of 
diagrams up to the order $\chi^{3/2}$, with $\chi$ the ratio
of the pion to the nucleon mass.
The net effect of all delta contributions at this order to the 
pion-deuteron scattering length is $\delta a^{\Delta}_{\pi d}=
(2.4 \pm 0.4)\times 10^{-3} \ m_\pi^{-1}$.
\end{abstract}

\section{Introduction}
\label{intro}

Chiral perturbation theory (ChPT) is the effective field theory of the
Standard Model at low energies allowing  for high accuracy calculations
of hadronic observables. It is a systematic expansion
around the chiral limit (vanishing quark masses) and vanishing
external momenta. ChPT can be applied to systems containing pions,
nucleons and external sources.
Here we focus on the $\pi NN$ system --- in particular
 the pion--deuteron system at threshold.

In the original formulation, only pions and nucleons appear
as dynamical degrees of freedom~\cite{Weinberg,ulfs}, whereas the impact
of baryon resonances as well as heavier mesons is absorbed into
certain low-energy constants. 
From phenomenological studies it is well known that the delta isobar $\Delta(1232)$
plays a very special role in low energy nuclear
dynamics~\cite{ericsonweise} as a consequence of the
relatively large $\pi N\Delta$ coupling and the
 quite small
delta--nucleon mass difference $\Delta=M_\Delta-M_N\simeq 3f_\pi$,
where $M_\Delta$, $M_N$, and $f_\pi$ denote the mass of the delta, of the
nucleon, and the pion decay constant, respectively.\footnote{Like the pion decay
constant, the delta--nucleon mass splitting does not vanish in the chiral
limit and thus this identification is more appropriate than  $\Delta\simeq
2m_\pi$, with $m_\pi$ the pion mass, as often found in the literature.}
In the effective field theory
sketched above this leads to unnaturally large values of some low--energy constants. 

It is also possible to include the delta as dynamical degree of
freedom in the effective field
theory~\cite{Jenkins:1991es,hemmert}. For the $\pi N$ system this
leads to a somewhat improved convergence of the chiral
expansion~\cite{nadiadelta}, however, no qualitative difference
appears compared to ChPT. In many cases, the representation of delta
effects through local pion-nucleon operators is quite accurate. As an
example we mention the successful analysis of threshold pion
photoproduction~\cite{photo}.  However, not only for energies of the
order of $\Delta$, but also at low energies in the spin sector the
explicit inclusion of the delta appears mandatory --- see, e.g., the
recent review~\cite{bernard}. It is, however, important to stress that
 the
delta-full theory in the single nucleon sector
 features more counter terms at a given order than
ChPT and has been much less systematically applied to low-energy
reactions.
\begin{figure}[t!]
\begin{center}
\epsfig{file=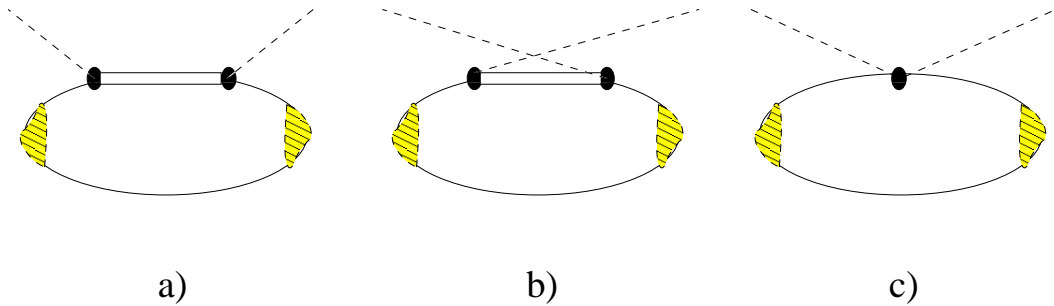, height=4cm, angle=0}
\caption{Classes of one-body diagrams that contribute to $\pi d$ scattering. Diagrams a) and b)
represent the one-body operators with the delta, diagram c) shows the corresponding
contact interaction in the delta-less theory.
Dashed lines denote pions and single (double) solid lines denote
nucleons (deltas). Solid black dots stand for interactions, whereas the hatched area shows
  the deuteron wave function.}
\label{dia}
\end{center}
\end{figure}
\noindent

In the present paper we investigate the role of the delta isobar in
the reaction $\pi d\to \pi d$ at threshold in chiral effective field
theory.  The reason why the explicit inclusion of the delta in pion
reactions on the two--nucleon system is beneficial is that the $\pi N$
amplitudes appear in the boosted frame due to the Fermi motion of the
nucleons. Let us for simplicity focus on the one--body terms with the
delta that contribute to $\pi d$ scattering at threshold (see
Fig.~\ref{dia}a) and b)). Then the $\pi N\Delta$ transition vertex is
linear in the nucleon momentum, $\vec p$, and the corresponding
embedded $\pi N\to\pi N$ transition potential is proportional to $\vec
p\, ^2$ times the
nucleon-delta propagator. The latter behaves as $1/(m_\pi-\Delta-\vec
p\, ^2/M_N)$ --- we point out that the width of the delta
is suppressed by two powers in the pion mass and thus
 does not
contribute to the order we are working.
  For static deltas, this propagator reduces to
$1/(m_{\pi}-\Delta)$ and the sum of diagrams a) and b) of
Fig.~\ref{dia} collapses to diagram c). Thus, in the latter case the
transition operator behaves like $\vec p\,^2$, whereas in the former
it approaches a constant for momenta larger than $|\vec p_\Delta \,
|\sim \sqrt{(\Delta-m_\pi)M_N}\sim 2.7m_\pi$ with the effect that the
static amplitude is more sensitive to the short range part of the
deuteron wave function and must be balanced by appropriate counter
terms, eventually of unnatural size.  The operator with the dynamical
delta, on the other hand, does not share this problematic
property. This point will be discussed in detail below.  The value of
$p_\Delta$ is numerically very close to $p_\mathrm{thr}=
\sqrt{M_Nm_\pi}\sim 2.6m_\pi$ --- the minimum initial momentum for the
reaction $NN\to NN\pi$. Therefore, in what follows we will use
\begin{equation}
p_\Delta \sim p_\mathrm{thr} \gg m_\pi \ .
\label{order}
\end{equation}
It was shown in
Ref.~\cite{disp} that the so--called dispersive corrections to the
$\pi d$ scattering length are suppressed by a factor $\chi^{3/2}$ relative
to the leading two--body operator with two Weinberg--Tomozawa (WT)
vertices, 
where $\chi=m_\pi/M_N$. The corresponding power counting, confirmed numerically, treated explicitly
the scale ${p_\mathrm{thr}\gg m_\pi}$ in line with the counting rules
for $NN\to NN\pi$~\cite{nnpi}.
The counting rule Eq.~(\ref{order}) automatically puts
the delta contributions in the same order as the dispersive
corrections, as we demonstrate below. 
There is one more class of contributions that can
scale as $\sqrt{\chi}$ in few-nucleon systems,
 namely the effect of $\pi NN$ cuts. However,
their impact on $\pi d$ scattering is negligible as
shown in Ref.~\cite{recoil}.
Thus, with this paper we complete 
the calculation  of diagrams at order $\chi^{3/2}$.

\medskip\noindent
The paper is structured as follows: in the next section we
describe the power counting. Results for the $\pi d$
scattering length  are reported
and compared to previous works in section \ref{sec:pid}.
The paper ends with some concluding remarks.

\section{Power counting}
\label{sec:powercount}

\begin{figure}[t!]
\begin{center}
\epsfig{file=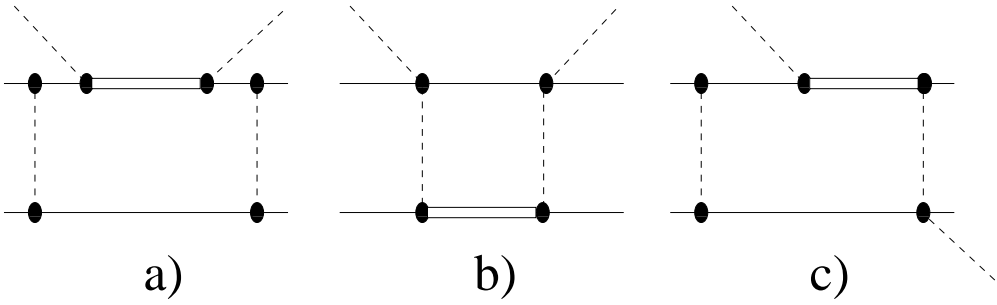, height=4cm, angle=0}
\caption{Typical subamplitudes that
contribute to $\pi NN$ scattering including
deltas at order $\chi^{3/2}$.}
\label{power}
\end{center}
\end{figure}

First of all we would like to remind the reader that the leading order
two--body operator with two WT vertices scales
simply as $m_{\pi}^2/(f_{\pi}^4p^2)\sim 1/f_{\pi}^4$ in the Weinberg
counting scheme, where for this diagram $p\sim m_\pi$.  
Below we will follow the logic of Ref.~\cite{disp}
and compare diagrams with the delta with this leading amplitude.
Let us start with the one--body terms depicted in Fig.~\ref{dia}a).
The transition amplitude scales as  
\be
A_{\pi N}^\Delta=
\frac1{f_\pi^2}\left(\frac{m_\pi}{M_N}\right)^2\frac{\vec p\, ^2}{m_\pi-\Delta-\vec p\, ^2/M_N+i\epsilon}.
\label{wpind}
\ee
As outlined in the Introduction, due to the squared
momentum in the numerator, momenta of order $|\vec p_\Delta \, |\sim
\sqrt{(\Delta-m_\pi)M_N}$ contribute to the full matrix element.
Therefore the nucleon before (after) the pion absorption (emission) is
off its mass shell by $ -{\vec p_\Delta}^{\,2}/2M_N \,$, i.e. by  about $m_\pi$.  
On the other hand, only on--shell amplitudes
are physically meaningful and should be compared to each other.  To
find the corresponding chiral order we should therefore estimate the
one loop diagram as shown in Fig.~\ref{power}a)\footnote{Note that the external 
nucleons in Fig.~\ref{power}a)
can also be off--shell, when the transition operators are 
convoluted with the
external wave functions. However, it is the central assumption of the
power counting that the
corresponding matrix element is dominated by (near) on--shell 
kinematics for these nucleons.}.
The estimate for
this diagram gives
\begin{equation} \hspace*{-0.7cm}\left[\left.\left(\frac{p^2}{f_\pi^2p^2}\right)^2\left(\frac{M_N}{p^2}\right)^2 A_{\pi N}^\Delta
\left(\frac{p^3}{(4\pi)^2}\right)\right]\right/\left(\frac1{f_\pi^4}\right) \sim \left\{\begin{array}{ll}
\mathcal{O}\left(\chi^2\right) & \mbox{for} \, p\sim m_\pi \\
\mathcal{O}\left(\chi^\frac32\right) & \mbox{for} \, p\sim
p_\Delta \\
\end{array}
\right.
\label{count}
\end{equation}
where the factors stand for the quantitative estimates for the two
one--pion exchange potentials, the two two--nucleon propagators,
the $\pi N\to\pi N$ transition potential through the delta, as defined 
in Eq.~(\ref{wpind}),
and the integral measure, consecutively. We stress that 
we do the power counting based on the expressions for time--ordered perturbation
theory, since we later work within this formalism. For details on this
we refer to Appendix E of Ref.~\cite{report}.
In the relation (\ref{count}) we estimated the contribution
of diagram \ref{power}a) for two regimes of pion momenta,
namely $p\sim m_\pi$ and $p\sim p_\Delta$.
For the identification of the chiral order
we used $M_N\sim 4\pi f_\pi$. We thus conclude that the power 
counting yields that
the dominant contribution of the delta loops is expected
to come from loop momenta of the order of $p_\Delta$, as argued
above.
Therefore the considered delta diagram contributes to the
same order as the so--called dispersive corrections to the $\pi d$
scattering length as discussed in Ref.~\cite{disp}.

The power counting for the other $\pi d$ diagrams goes in just the
same way.  E.g., for the diagrams b) and c) of Fig.~\ref{power} we
also find the chiral order $\chi^{3/2}$. The power counting for a
diagram of type b), however with a two--nucleon intermediate state, was
discussed in detail in Ref.~\cite{disp}. Since we count the
nucleon-delta propagator in the same manner as the two--nucleon
propagator, i.e. as $1/m_\pi$, it becomes obvious that the
corresponding diagrams contribute at the same order.  In the
estimation of the chiral order of diagrams b) and c) we used that the
leading $\pi N\to \pi N$ vertex scales as $m_\pi$ and not as the pion
energy, regardless of the relatively large momentum running in the
loop.  The terms dropped are higher order in the chiral expansion.
This is in line with the findings of Ref.~\cite{lensky2} for the
reaction $NN\to d\pi$.
\begin{figure}[t!]
\begin{center}
\epsfig{file=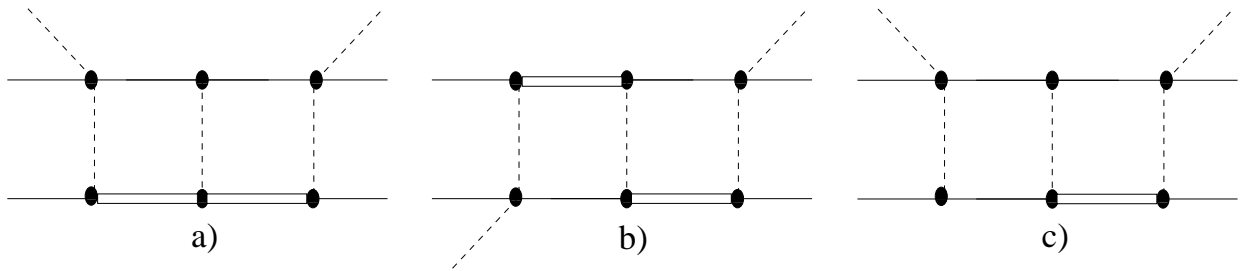, height=3.5cm, angle=0}
\caption{Typical subamplitudes that
contribute to $\pi NN$ scattering including
deltas at order $\chi^2$. These (and all the others of such kind) are not included
in this work.}
\label{power2}
\end{center}
\end{figure}
\noindent
Every additional loop including deltas leads at least to an
additional factor $p^3/(4\pi)^2  \times\,1/m_\pi \, \times 1/f_\pi^2$
for the integral measure, the $N\Delta$ propagator and 
the leading $N\Delta$ interaction, consecutively. Therefore, diagrams with
an intermediate $N\Delta\to N\Delta$ transition, as shown
in Fig.~\ref{power2}a) and b), and those with an
intermediate $NN\to N\Delta$ transition, diagram c), are suppressed
by one power in $p/M_N\sim\chi^{1/2}$ compared to the diagrams 
shown in Fig.~\ref{power} and will not be considered
in this work.  
Consequently, from the naive power counting arguments we can expect the
leading delta contribution to be of order of
$(m_{\pi}/M_N)^{3/2}\,|a_{\pi d}^\mathrm{double}|\simeq 0.06\, |a_{\pi
d}^\mathrm{exp}|\simeq 1.6\times 10^{-3} m_{\pi}^{-1}$ where we used that
$|a_{\pi d}^\mathrm{exp}|\simeq26\times 10^{-3} m_{\pi}^{-1}$ and that the
real part of the scattering length is dominated by the double
rescattering term with two WT vertices --- giving rise to $a_{\pi
d}^\mathrm{double}$. This estimation is fully in line with our numerical
results as given in the next section.
In addition, as stated already, we do not consider
terms of order $\chi^2$. Using the same reasoning as above, we
can also estimate the theoretical uncertainty of our calculation
as 
\begin{equation}
\Delta a^{\rm theor} = (m_{\pi}/M_N)^{2}\,|a_{\pi d}^\mathrm{double}|\simeq 0.6\times 10^{-3} m_{\pi}^{-1} \ .
\label{theoun}
\end{equation}
At order $\chi^2$ also the leading $NN\pi\to NN\pi$ counter term
contributes to $\pi d$ scattering with up--to--now unknown
coefficient. Therefore $\Delta a^{\rm theor}$ represents at the same
time an estimate for the theoretical accuracy for the extraction
of the isoscalar scattering length $a^+$ from $\pi d$
scattering~\cite{indien}.  For a further improvement in the accuracy
of the calculation, input from other reactions is needed to fix the
value of the counter term. One possible source of this information
could be the reaction $NN\to NN\pi^0\pi^0$.

\section{Results and comparison to previous works}
\label{sec:pid}

\begin{table}[p]
\begin{center}
\caption{Delta contributions to the real part of $a_{\pi d}$ in units
of $m_\pi^{-1} \times 10^{-3}$ calculated with $h_A=2.77$. The results are shown for the 
phenomenological $NN$ potentials Paris~\cite{paris}, AV18~\cite{argonne},
CD-Bonn~\cite{cdbonn} (CDB), and CCF~\cite{ccf} as well as for the wave functions
of the N$^2$LO chiral $NN$ interaction~\cite{evgeni} based on the pairs
of regulators $\{600, 500\}$ (EGM1), $\{550, 600\}$ (EGM2) and $\{450, 700\}$ (EGM3). 
All integrals are evaluated up to 1 GeV -- 
the contributions of higher momenta are negligible.}
\vskip 0.33cm
\begin{tabular}{|c| c c |p{1.cm}|p{1.cm}|p{1.cm}||p{1.cm}||p{1.cm}|p{1.cm}|p{1.cm}|}
\hline
& & & Paris &AV18&CDB&CCF   & EGM1 & EGM2 & EGM3\\
\hline
$e1$ & \parbox[c]{3cm}{\epsfig{file=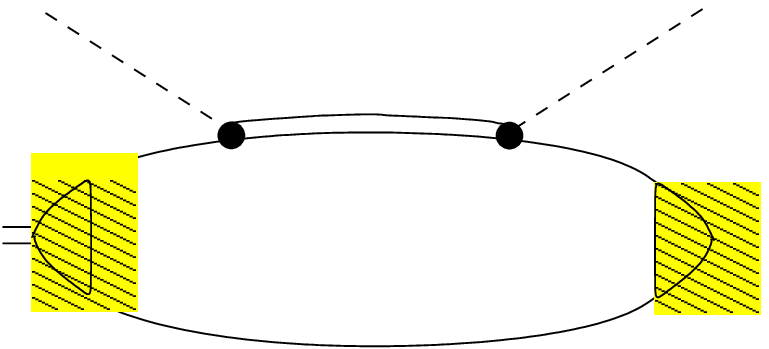, height=1.6cm,width=2.6cm}}  &              $=$ &  $+1.89$   &  $+1.92$  & $+1.77$ &  $+1.81$              & $+1.69$ & $+1.66$ & $+1.57$            \\ 
$e2$ & \parbox[c]{3cm}{\epsfig{file=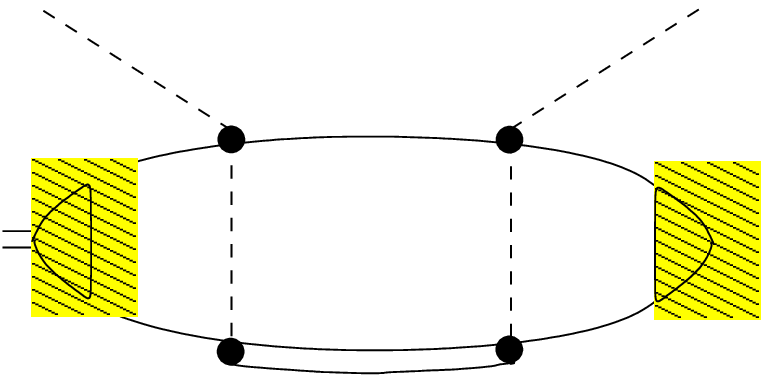, height=1.6cm,width=2.6cm}}  &              $=$ &  $+0.54$   &  $+0.55$  & $+0.73$ &  $+0.56$             & $+0.72$ & $+0.79$ & $+0.84$             \\ 
$e3$ & \parbox[c]{3cm}{\epsfig{file=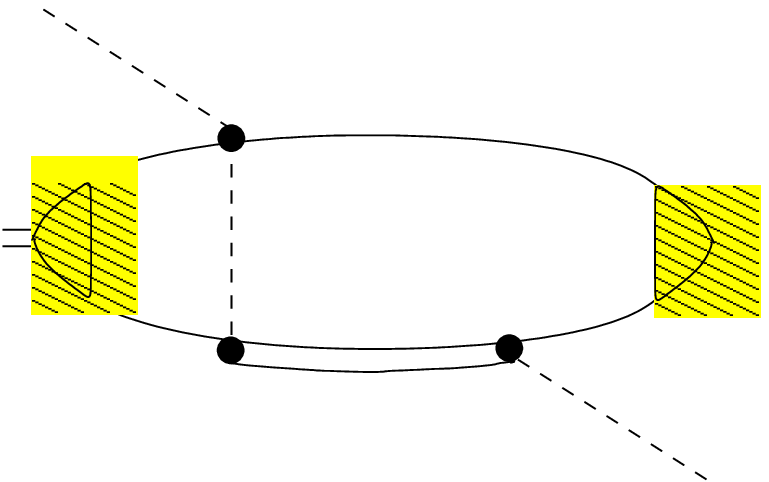, height=2.0cm,width=2.6cm}}&              $=$ & $-0.70$    &  $-0.73$  & $-0.94$& $-0.72$              & $-0.91$ & $-1.02$ & $-1.08$            \\
\hline 
& & & & & & & & & \\
& sum of this group &                                                                        $=$ & $+1.73$    &  $+1.74$  & $+1.56$ & $+1.65$              & $+1.50$ & $+1.43$ & $+1.33$              \\ 
& & & & & & &  & & \\
\hline
\hline
$f1$ &  \parbox[c]{3cm}{\epsfig{file=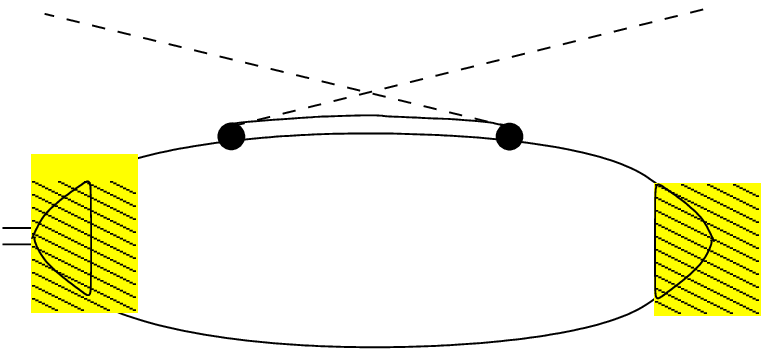, height=1.6cm,width=2.6cm}} &         $=$ & $+0.84$    & $+0.85$   & $+0.75$ & $+0.79$              & $+0.69$ & $+0.67$ & $+0.63$           \\ 
$f2$ & \parbox[c]{3cm}{\epsfig{file=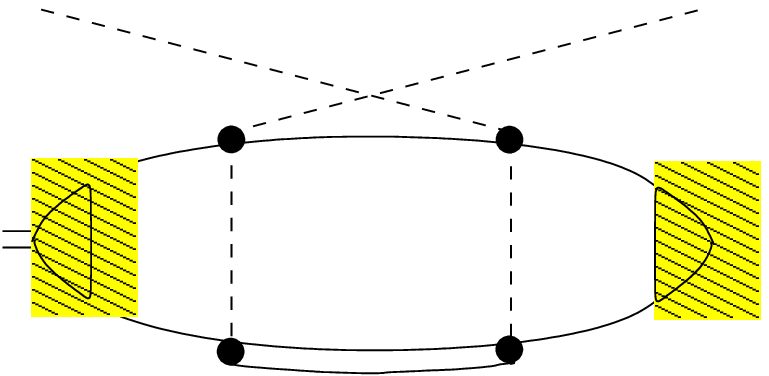, height=1.6cm,width=2.6cm}} &          $=$ & $+0.13$    & $+0.14$   & $+0.21$ & $+0.14$             & $+0.21$  & $+0.24$  & $+0.26$            \\ 
$f3$ & \parbox[c]{3cm}{\epsfig{file=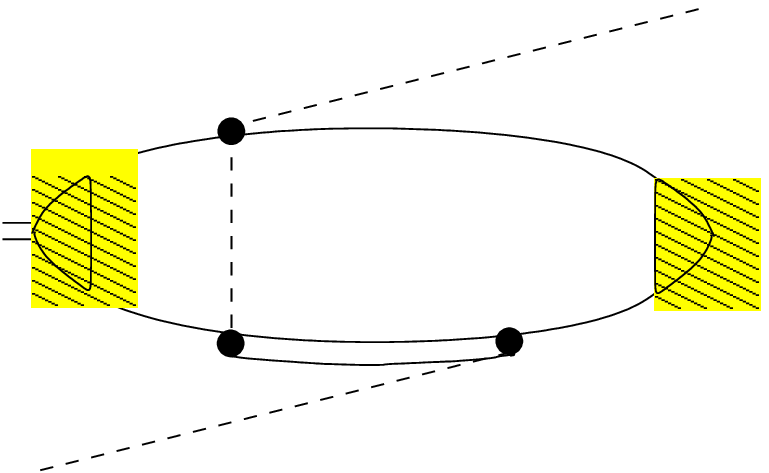, height=2.0cm,width=2.6cm}} &        $=$ & $-0.05$    & $-0.05$   & $-0.14$& $-0.06$             & $-0.14$  & $-0.19$  & $-0.22$            \\
\hline 
& & & & & & &  & & \\
& sum of this group &                                                                        $=$ & $+0.92$    & $+0.94$   & $+0.82$ & $+0.87$              & $+0.76$ & $+0.72$ & $+0.67$           \\ 
& & & & & & &  & & \\
\hline 
\hline 
& & & & & & &  & & \\
& total sum  & $=$ &
 $+2.65$    & $+2.68$   & $+2.38$   &  $+2.52$& $+2.26$ & $+2.15$ & $+2.00$\\ 
& & & & & & &  & & \\
\hline
\end{tabular}
\label{results}
\vspace{0.3cm}
\end{center}
\end{table}

Although the vertex structure we use for the $\pi N\Delta$ vertex is
standard (see, e.g., Ref.~\cite{Rocha} and references therein), in order to
keep the paper self-contained and to
fix the normalization we present it here (note that our
vertex normalization differs by a factor of two
compared e.g.  to the one of Ref.~\cite{evgeninew}):
\begin{eqnarray}
\nonumber
 {\cal L}^{(0)} &=  &
          \frac{h_{A}}{2 f_{\pi}}[N^{\dagger}(\boldT\cdot
          \vec{S}\cdot\vec{\nabla}{\mathbf{\boldpi}})\Psi_\Delta +\mathrm{h.c.}] \ ,\\
{\cal L}^{(1)} &=& {-}\frac{h_{A}}{
        2 M_\Delta f_{\pi}}[
        iN^{\dagger}\boldT\cdot\dot{\boldpi}\vec{S}\cdot\vec{\nabla}
        \Psi_\Delta {+} \mathrm{h.c.}] \ .
\end{eqnarray}
Here $h_A$ denotes the leading $\Delta N \pi$ coupling, and $\vec S$ and $\bold T$ are
the spin and isospin transition matrices, normalized such that
\begin{eqnarray}
\nonumber
S_iS_j^\dagger &=& 
\frac{1}{3}(2\delta_{ij}-i\epsilon_{ijk}\sigma_k) \ , \\
T_iT_j^\dagger &=&
 \frac{1}{3}(2\delta_{ij}-i\epsilon_{ijk}\tau_k) \ .
\end{eqnarray}
In our calculations we use $f_\pi=92.4$ MeV and 
$h_A= 3g_A/\sqrt{2}\simeq 2.1g_A=2.77$,
where $g_A=1.32$ is the axial--vector coupling of the nucleon (derived from
the Goldberger-Treiman relation). The relation between $h_A$ and $g_A$ 
can be derived from large $N_c$ arguments and the resulting
coupling gives a reasonable description of the delta width at tree
level~\cite{norbertsdelta}.
Very similar values were shown to be consistent with the $\pi N$
phase shifts in the delta region~\cite{ellisdelta,danieldelta}.
 It should be noted, however, that the dispersion theoretical
analysis of Ref.~\cite{hoehlerdelta} leads to the considerably
lower value of $h_A = 2.1$.


In Table \ref{results} we show the results 
of our numerical calculations for the complete
set of diagrams with the delta isobar that contribute at order
$\chi^{3/2}$.
These numbers were produced using our prefered value
$h_A=2.77$.
 In order
to study the model dependence of the results
we performed the calculations for various
$NN$ potentials. Note that we used phenomenological $NN$ models 
without \cite{paris,argonne,cdbonn} and with \cite{ccf} explicit delta degree of freedom, 
as well as three variants of $NN$ wave functions derived within
chiral effective field theory \cite{evgeni}. We remark that ideally one would also
use chiral wave functions with explicit deltas. However,
up to now corresponding wave functions of sufficient accuracy exist only
for higher partial waves~\cite{evgeninew}. Using the different potentials
mentioned, we obtain
\begin{equation}
\delta a_{\pi d}^\Delta = (2.38\pm 0.40) \times
\,10^{-3} \ m_\pi^{-1} \ ,
\label{result}
\end{equation}
where the central value is the arithmetic
average of the results for the seven different potentials
and the uncertainty reflects the variations in the
results. Consistency of the power counting 
demands that the dependence on the $NN$ potential
used does not exceed the contribution estimated
for the leading counter term, $\Delta a^{\rm theor}$ given
in Eq.~(\ref{theoun}), that can absorb this
dependence. In this sense  Eq.~(\ref{result}) is
an additional confirmation for the
consistency of the power counting employed.

All diagrams evaluated contain the $\pi N\Delta$
coupling constant $h_A$ squared. Thus, to see
the impact of a value as low as $h_A=2.1$ on
our results, the numbers given in Table~\ref{results}
simply need to be rescaled. We then would get
$
\left.\delta a_{\pi d}^\Delta\right|_{h_A=2.1}=(1.4\pm 0.2) \times
\,10^{-3} \ m_\pi^{-1}.
$ 
However, we regard Eq.~(\ref{result}) as our
main result, since the value employed for $h_A$
can be extracted from fits to the $\pi N$ system
in the delta region based on calculations
consistent with the one discussed here~\cite{ellisdelta,danieldelta}.

Note that the results from the chiral
wave functions are systematically lower than those
from the phenomenological potentials, 
which might be a consequence of differences
of the $NN$ interactions at intermediate range.
This finding does not come unexpected. However,
calculations to higher orders are necessary
to draw more firm conclusions.

The results we found depend only very weakly on the $NN$ models used.
In contrast to this, many previous works find a significant model
dependence when using phenomenological parameterizations for some of
the diagrams discussed above.  For example, in
Refs.~\cite{BK,doeringoset} diagram $e3$ of Table \ref{results} was
included by replacing the delta propagator and vertices by the
phenomenological $\pi N$ $p$--wave amplitude expressed in terms of the
$p$-wave volumes $c_0$ and $c_1$ and evaluated in the boosted frame
(this is called SP-interference term in Ref.~\cite{BK}). 
The evaluated
matrix element shows a significant model dependence for it scales with
the deuteron wave function at the origin (for a more detailed study of
the model dependence of this quantity see Ref.~\cite{mitandreas}). To
illustrate how large the model dependence of the corresponding
amplitude could be, we give in Table~\ref{phenom} the results for the
diagram $e3$ calculated with the static $\Delta$ propagator. The
results vary by more than a factor of four when different $NN$ models are
employed. The corresponding results for the diagram $e1$ (see Refs.~\cite{BK,doeringoset,Erics} 
for the corresponding phenomenological calculations), 
also given in Table~\ref{phenom}, differ by a factor 1.6.  
 As stressed already in the Introduction, once the
kinetic energy of the delta is kept in the propagator, as demanded by
the power counting, the above problem disappears and almost 
model-independent results emerge (see lines $e1$ and $e3$ in Table \ref{results}).
\begin{table}[t]
\begin{center}
\caption{Results for diagrams $e1$ and $e3$ evaluated with the static  delta propagator in units 
of $ 10^{-3} m_\pi^{-1}$. Integrals are evaluated up to 1 GeV.}
\vskip 0.33cm
\begin{tabular}{|c| c   |p{1.cm}|p{1.cm}|p{1.cm}||p{1.cm}||p{1.cm}|p{1.cm}|p{1.cm}|}
\hline
& &   Paris &AV18& CDB &CCF & EGM1 &EGM2 &EGM3\\
\hline
$e1$ & \parbox[c]{3cm}{\epsfig{file=Deltadir.eps, height=1.6cm,width=2.6cm}} &   $+3.0$   &  $+3.1$  & $+2.5$ &  $+2.8$              & $+2.2$ & $+2.1$ & $+1.9$            \\ 
\cline{3-9}
\hline
 $e3$ & \parbox[c]{3cm}{\epsfig{file=DeltaWTdir.eps, height=2.0cm,width=2.6cm}}&  $-0.3$   &  $-0.4$  & $-0.8$ &  $-0.4$             & $-1.0$ & $-1.3$ & $-1.4$            \\ 
\cline{3-9}
\hline
\end{tabular}
\label{phenom}
\vspace{0.3cm}
\end{center}
\end{table}
\noindent
In an effective field theory calculation without explicit deltas,
diagrams $e1$ and $f1$ were included effectively as  so--called
boost corrections~\cite{BBEMP}. The resulting contribution to the $\pi
d$ scattering length turned out to be quite sizable, namely $(3 - 5)
\times 10^{-3} \ m_\pi^{-1}$, depending on the regulator used for the
$NN$ potential. Evidently, the spread in the results is well above the
estimate of Eq.~(\ref{theoun}), which, again, is a consequence of
dropping the kinetic energy of the delta isobar. In the theory without
deltas the pertinent one--body operator scales with the square of the
nucleon momentum and therefore the corresponding expectation value is
proportional to the nucleon kinetic energy inside the deuteron ---
this quantity is strongly model-dependent~\cite{mitandreas}, which
indicates that the power counting in the delta-less theory requires
further modification. However, the boost term (see Ref.~\cite{BBEMP})
is proportional to the low energy constant $c_2$, which is known to be
largely saturated by the delta isobar \cite{BKM1,evgeninew}.  In the
analysis of the $\pi N$ system \cite{BKM1} it was shown that the
explicit evaluation of the leading order delta contribution results in
a reduction of the value of $c_2$ from about 3.3~GeV$^{-1}$ to about
$0.5$~GeV$^{-1}$.  In the very recent analysis of the $NN$ system
including explicitly the delta at NLO~\cite{evgeninew}, an analysis of
$\pi N$ threshold coefficients was performed. Given the parameters
utilized there, the value of $c_2$ is reduced to $-0.25$ GeV$^{-1}$.
A reduction of the $\pi N\Delta$ coupling by 30\% as demanded by a
dispersive analysis of the resonance contribution to the pion-nucleon
$P_{33}$ phase shifts~\cite{hoehlerdelta} leads to a reduced $c_2 =
0.83\,$GeV$^{-1}$. All these values are consistent within the
uncertainty of the various contributions to the low-energy constants
given in Ref.~\cite{BKM1}.  Therefore the value of $c_2$ is reduced by
a large factor once the delta contribution is taken out. We have
calculated the residual boost correction using the expressions given
in Ref.~\cite{BBEMP} with $\mathrm{N^2LO}$ chiral wave functions and
with $c_2=-0.25$ GeV$^{-1}$ and found it to be as small as $-(5.7 \ldots
6.6) \times \,10^{-4} \ m_\pi^{-1}$.  Consequently, this correction is
of the same size as the estimated uncertainty of the calculation
(see. Eq.~(\ref{theoun})) and thus does not contribute significantly
anymore.

It should be stressed that it is not compulsory for a consistent
calculation of the $\pi d$ scattering length that the delta is
included explicitly. Also a calculation without deltas is obviously
equally justified. As usual the effects of the delta would then be
parameterized by local counter terms of the type $\pi NN\to \pi NN$
with up-to-now unknown coefficients. The conclusion to be drawn from
our studies is that in order to perform calculations with the accuracy
of the order of the uncertainty estimate given in Eq.~(\ref{theoun})
it is necessary to include a dynamical delta, as long as no additional
information on the size of the counter term is available. On the other
hand, for a consistent inclusion of isospin breaking effects, that are
known to be important \cite{mrr_iso}, more theoretical work on the
treatment of effects from quark masses and virtual photons in the
delta-full theory would be useful.

\section{Conclusions}
\label{sec:con}

In this work we calculated the leading contributions of the
$\Delta(1232)$ to the $\pi d$ scattering length in effective field
theory. As expected, inclusion of the delta leads to an improved
convergence for the isospin-symmetric operators that contribute to
this reaction.  We have also compared our results to other approaches
and discussed the differences.

In the power counting employed the delta starts to 
contribute at order $\chi^{3/2}$, relative to
the leading two--nucleon contribution, given
by two subsequent $\pi N$ scatterings on the two
different nucleons. At the same chiral order 
the so--called dispersive corrections evaluated
in Ref.~\cite{disp} contribute as well, and
with this work we complete the evaluation
of diagrams at that order. In Ref.~\cite{disp} the dispersive
corrections were evaluated for a particular $NN$
potential. 
When repeating the calculation with the four different phenomenological 
$NN$ potentials employed in the present study (note: the
chiral wave functions could not be used here, since
for the dispersive corrections the wave functions are needed
also at pion production threshold, where the chiral
wave functions are not applicable anymore) we find
\begin{equation}
\delta a_{\pi d}^{\rm disp} = (-2.9\pm 1.4) \times
\,10^{-3} \ m_\pi^{-1} \ ,
\label{result_disp}
\end{equation}
where the first number is the mean value for
the various potentials and the second number
reflects the theoretical uncertainty of this
calculation estimated conservatively --- see Ref.~\cite{disp} for details. The variation of the results for the different
potentials lies well within this uncertainty band.
Note, that the uncertainty can be reduced by a calculation
of $NN\to d\pi$ to next--to--next--to--leading order,
which is planned for the near future.
We therefore find for the total contribution at
order $\chi^{3/2}$
\begin{equation}
\delta a_{\pi d}^\Delta+\delta a_{\pi d}^{\rm disp}=(-0.6\pm 1.5) \times
\,10^{-3} \ m_\pi^{-1} \ ,
\end{equation}
where we added the uncertainties given in Eqs.~(\ref{result}) and (\ref{result_disp})
in quadrature.
Thus, we conclude that the net effect of the diagrams
that contribute at order $\chi^{3/2}$ is very small.
Note that the occurring cancellation is accidental because 
very different physics contributes to the
two classes of diagrams.

One important consequence of our investigations is
that once the delta isobar is treated dynamically, as it is done in this paper,
the so--called boost corrections give rise  to an insignificant contribution
in the theoretical analysis of the $\pi d$ scattering
length. Furthermore, for the same reason the phenomenological inclusion of
pion rescattering (the so--called SP interference term) through a boosted $p$--wave
amplitude, used in Refs.~\cite{BK,doeringoset}, is expected to yield a very
small contribution, well within the theoretical uncertainty given here --- see
also the corresponding discussion in Ref.~\cite{disp}.

With this work all strong, isospin--symmetric contributions
to the $\pi d$ scattering length have been calculated to very high
accuracy. In principle we could now extract the isoscalar
$\pi N$ scattering length, $a^+$, directly from the $\pi d$ scattering
length, since
\begin{equation}
a_{\pi^- d}= 2a^++\left<\mbox{few--body corrections }(a^-)\right> \ ,
\label{whatweget}
\end{equation}
where $a^-$ denotes the isovector scattering length.  In this
expression additional terms that contain $a^+$ were neglected for they
are numerically negligible.  However, in addition isospin violating
effects are known to be quite sizable. Therefore, in
Eq.~(\ref{whatweget}) we should replace $2a^+$ by $a_{\pi^-
p}+a_{\pi^- n}$ which agrees to the former only, if isospin were
an exact symmetry.  Furthermore, few--body corrections involving
virtual photons, in addition to those calculated in Ref.~\cite{disp},
are potentially important.  For the $\pi^- d$ system so far only the
leading isospin violating corrections were evaluated~\cite{mrr_iso}.
To this order the largest theoretical uncertainty emerged from the
appearance of the low--energy constants $f_1$ and $c_1$. It is
intriguing to observe, however, that those appear in the same linear
combination in both $a_{\pi^-p}$ and $a_{\pi^-n}$. Thus, one is in the
position to extract $a_{\pi^-p}+a_{\pi^-n}$ with high accuracy from a
combined analysis of pionic deuterium and pionic hydrogen even without
detailed knowledge on $f_1$ (see also Ref.~\cite{trento}). However, it
remains to be seen if the corrections at next--to--leading order in
isospin violation do not distort this picture. Corrections at this
order for the $\pi^- p$ system were evaluated in
Refs.~\cite{piploop,nadia2} and turned out to be quite sizable,
especially those that come from the pion mass difference.  In order to
push also the calculation for the $\pi d$ system to a similar level of
accuracy in isospin violation, the $\pi^- n$ scattering amplitude as
well as some virtual photon exchanges in the $\pi^- d$ system are
still to be calculated.

\vspace{0.5cm}

\noindent 
{\bf Acknowledgments}

\noindent 
We thank Evgeny Epelbaum, Andreas Nogga, Daniel Phillips, and Akaki Rusetsky for useful discussions.
This research is part of the EU Integrated Infrastructure Initiative
Hadron Physics Project under contract number RII3-CT-2004-506078, and
was supported also by the DFG-RFBR grant no. 05-02-04012 (436 RUS
113/820/0-1(R)) and the DFG SFB/TR 16 "Subnuclear Structure of Matter".  A.~K. and
V.~B. acknowledge the support of the Federal Agency of Atomic Research
of the
Russian Federation.

\renewcommand{\theequation}{A.\arabic{equation}}
\section*{Appendix}  
\setcounter{equation}{0}  

In this appendix we present the explicit expressions for the
amplitudes given in Table~\ref{results}.  Note that in accordance with the power
counting, we only keep those amplitudes that contain intermediate states with the nucleon, the delta
and at most real pions. The calculation is done in time--ordered perturbation
theory (TOPT). Especially, we  dropped the so--called stretched boxes. 
The corresponding correction to the $\pi d$ scattering length due to the delta isobar is 
\be
\delta  a_{\pi d}^\Delta= a_{\pi d}^\Delta(q_0=m_{\pi})+a_{\pi d}^\Delta(q_0=-m_{\pi})
\ee
where the first and second terms correspond to the direct and crossed diagrams
of  Table~\ref{results}, respectively.
Here 
\be
 {a_{\pi d}^{\Delta}}(q_0)&=& -\frac{h_A^2m_{\pi}^2 }{48\pi f_{\pi}^6\, (1{+}m_{\pi}/2M_N)}
\int \frac{d^3q}{(2\pi)^3}
\frac{q^2}{q_0-\Delta -\displaystyle q^2 / 2M_{N\Delta}}\; 
\left(I_{^3P_1}+I_{^5P_1}+I_{^5F_1} \right)
\label{explexpr}
\ee
where $M_{N\Delta}=M_NM_{\Delta}/(M_N+M_\Delta)$ and $I_{^{2S+1}L_J}$ are the partial wave amplitudes squared 
that correspond to the  decomposed
intermediate $N\Delta$ state
\be
\nonumber
I_{^3P_1}&=& \frac{1}{9}  \left[I_1^{\Delta}(q)-\frac{3}{2\sqrt{2}}I_2^{\Delta}(q) 
-\frac{2 f_{\pi}^2}{M_{\Delta}} \left (u(q)+\frac{w(q)}{\sqrt{2}}\right) \right]^2~,  \\
I_{^5P_1}&=& \frac{5}{9} \left[I_1^{\Delta}(q)-\frac{3}{10\sqrt{2}}I_3^{\Delta}(q) 
-\frac{2 f_{\pi}^2}{M_{\Delta}} \left (u(q)-\frac{w(q)}{5\sqrt{2}}\right)\right]^2~,\label{I_PW}\\
\nonumber
I_{^5F_1}&=& \frac{3}{5}  \left[I_4^{\Delta}(q)-\frac{2 f_{\pi}^2}{M_{\Delta}} w(q) \right]^2 ~, 
\ee
with the $I_i^{\Delta}$ denoting the integrals that correspond to the overlap of 
the deuteron wave function ($u(q)$ and $w(q)$ for the S- and D-waves,  respectively) with 
the one-pion-exchange operator
\be
\nonumber
I_1^{\Delta}(q)&=&-\int \frac{d^3p}{(2\pi)^3} \frac{1+({\vec p}\cdot{\vec q}\,)/q^2}{2\omega_{{\vec p}
+{\vec q}}} \left (u(p)+\frac{w(p)}{\sqrt{2}}\right)\left(\frac{1}{P_1} + \frac{1}{P_2^{\Delta}}\right),\\
I_2^{\Delta}(q)&=&-\int \frac{d^3p}{(2\pi)^3} 
\frac{1-({\vec p}\cdot{\vec q}\,)^2/(p^2q^2)}
{2\omega_{{\vec p}+{\vec q}}} w(p)
\left(\frac{1}{P_1} + \frac{1}{P_2^{\Delta}}\right)~,\\\nonumber
I_3^{\Delta}(q)&=&-\int \frac{d^3p}{(2\pi)^3} 
\frac{3+4({\vec p}\cdot{\vec q}\,)/q^2+({\vec p}\cdot{\vec q}\,)^2/(p^2q^2)}
{2\omega_{{\vec p}+{\vec q}}} w(p)
\left(\frac{1}{P_1} + \frac{1}{P_2^{\Delta}}\right)~,
\\\nonumber
I_4^{\Delta}(q)&=&-\frac{1}{2}\int \frac{d^3p}{(2\pi)^3} 
\frac{-1-3({\vec p}\cdot{\vec q}\,)/q^2+3({\vec p}\cdot{\vec q}\,)^2/(p^2q^2)+
5({\vec p}\cdot{\vec q}\,)^3/(p^3q^3)}
{2\omega_{{\vec p}+{\vec q}}} w(p)
\left(\frac{1}{P_1} + \frac{1}{P_2^{\Delta}}\right)~.
\ee
Here $P_1$  and $P_2^{\Delta}$ correspond to the TOPT components of the pion propagator
\be\nonumber
P_1&=&q_0-\omega_{{\vec p}+{\vec q}}-(p^2+q^2)/2M_N, \\
P_2^{\Delta}&=&-\omega_{{\vec p}+{\vec q}}-\Delta -p^2/2M_N-q^2/2M_{\Delta}
\ee
with $\omega_{\vec q}=\sqrt{{\vec q}\:^2+m_{\pi}^2}$.
The diagrams of Table~\ref{results} can be easily matched to the
individual terms of Eqs.~(\ref{explexpr}) and (\ref{I_PW}): the very
last terms on the r.h.s. of each amplitude $I_{^{2S+1}L_J}$ in
Eqs.~(\ref{I_PW}), proportional to the deuteron wave functions squared,
correspond to the diagrams of type 1 ($e1$ and $f1$), type 2 contains
$I_i^{\Delta}$ amplitudes squared, whereas the interference terms of
type 3 contain the rest.  For the direct terms, labeled as $e$ in
Table~\ref{results}, one needs to take $q_0=m_\pi$ and for the crossed
terms, labeled as $f$ in that Table, $q_0=-m_\pi$.  Finally, we remark
that all integrals are evaluated up to a sharp momentum cut--off of 1
GeV. All higher momentum contributions are negligible and anyway are
to be absorbed in a counter term that is to be included at order
$\chi^2$. 
calculated with different wave functions demonstrates nice
convergence.


\begin{thebibliography}{99}
\bibitem{Weinberg} S. Weinberg, Physica A {\bf 96} (1979) 327.
\bibitem{ulfs}
V.~Bernard, N.~Kaiser and U.-G.~Mei\ss ner,
Int. \ J. \ Mod. \ Phys. \ E {\bf 4} (1995) 193  [arXiv:hep-ph/9501384]. 
\bibitem{ericsonweise}
T. Ericson und W. Weise, {\em Pions and Nuclei} (Clarendon Press, Oxford,
  1988).

\bibitem{Jenkins:1991es}
  E.~Jenkins and A.~V.~Manohar,
  Phys.\ Lett.\  B {\bf 259} (1991) 353.

\bibitem{hemmert}
T.R. Hemmert, B.R. Holstein, and J. Kambor, J. Phys. G {\bf 24}
(1998) 1831 [arXiv:hep-ph/9712496].
\bibitem{nadiadelta}
  N.~Fettes and U.-G.~Mei\ss ner,
  Nucl.\ Phys.\  A {\bf 679} (2001) 629
  [arXiv:hep-ph/0006299].
\bibitem{photo}
  V.~Bernard, N.~Kaiser and U.-G.~Mei{\ss}ner,
  Z.\ Phys.\  C {\bf 70} (1996) 483
  [arXiv:hep-ph/9411287].
\bibitem{bernard}
  V.~Bernard,
  arXiv:0706.0312 [hep-ph].
\bibitem{disp} V.~Lensky, V.~Baru, J.~Haidenbauer, C.~Hanhart,
   A.~E.~Kudryavtsev and U.-G.~Mei\ss ner, 
   Phys. Lett. B {\bf 648} (2007) 46 [arXiv:nucl-th/0608042].  
\bibitem{nnpi}
 C.~Hanhart, U.~van Kolck and G.~A.~Miller,
  Phys.\ Rev.\ Lett.\  {\bf 85} (2000) 2905
  [arXiv:nucl-th/0004033]; C.~Hanhart and N.~Kaiser,
  Phys.\ Rev.\  C {\bf 66}, 054005 (2002)
  [arXiv:nucl-th/0208050].

\bibitem{recoil}
  V.~Baru, C.~Hanhart, A.~E.~Kudryavtsev and U.-G.~Mei{\ss}ner,
  Phys.\ Lett.\  B {\bf 589} (2004) 118
  [arXiv:nucl-th/0402027].
\bibitem{report}
C.~Hanhart,
  Phys.\ Rept.\  {\bf 397} (2004) 155
  [arXiv:hep-ph/0311341].

\bibitem{lensky2}
  V.~Lensky, V.~Baru, J.~Haidenbauer, C.~Hanhart, A.~E.~Kudryavtsev and U.-G.~Mei\ss ner,
  Eur.\ Phys.\ J.\ A {\bf 27} (2006) 37
  [arXiv:nucl-th/0511054].
\bibitem{indien}
  C.~Hanhart,
  arXiv:nucl-th/0703028.
\bibitem{Rocha}
 C.~da Rocha, G.~Miller and U.~van Kolck,
  Phys.\ Rev.\  C {\bf 61} (2000) 034613 
  [arXiv:nucl-th/9904031].
\bibitem{evgeninew}
  H.~Krebs, E.~Epelbaum and U.-G.~Mei\ss ner,
  Eur. \ Phys. \ J. A {\bf 32} (2007) 127
  [arXiv:nucl-th/0703087].
\bibitem{norbertsdelta}
  N.~Kaiser, S.~Gerstendorfer and W.~Weise,
  Nucl.\ Phys.\  A {\bf 637} (1998) 395
  [arXiv:nucl-th/9802071].
\bibitem{ellisdelta}
  P.~J.~Ellis and H.~B.~Tang,
  Phys.\ Rev.\  C {\bf 56} (1997) 3363
  [arXiv:hep-ph/9609459].
\bibitem{danieldelta}
  V.~Pascalutsa and D.~R.~Phillips,
  Phys.\ Rev.\  C {\bf 67} (2003) 055202
  [arXiv:nucl-th/0212024].
\bibitem{hoehlerdelta}
  G. H\"ohler in Landolt--B\"ornstein, Vol. 9 b2, ed. H. Schopper (Springer, Berlin, 1983).
\bibitem{paris}   M. Lacombe et al., Phys. Rev. C {\bf 21} (1980) 861.
\bibitem{argonne}    R.~B.~Wiringa, V.~G.~J.~Stoks and R.~Schiavilla,
  Phys.\ Rev.\  C {\bf 51} (1995) 38
  [arXiv:nucl-th/9408016].             
\bibitem{cdbonn}
  R.~Machleidt, 
  Phys.\ Rev.\ C {\bf 63} (2001) 024001
  [arXiv:nucl-th/0006014].
\bibitem{ccf}  J. Haidenbauer, K. Holinde, and M.B. Johnson, 
Phys. Rev. C {\bf 48} (1993) 2190.
\bibitem{evgeni}  E.~Epelbaum, W.~Gl\"ockle and U.-G.~Mei\ss ner,
  Nucl.~Phys.~A {\bf 747}  (2005) 362 [arXiv:nucl-th/0405048].
 \bibitem{BK}
V.~V.~Baru and A.~E.~Kudryavtsev,
  Phys.\ Atom.\ Nucl.\  {\bf 60} (1997) 1475
  [Yad.\ Fiz.\  {\bf 60} (1997) 1620].
\bibitem{doeringoset}
  M.~D\"oring, E.~Oset and M.~J.~Vicente Vacas,
  Phys.\ Rev.\ C {\bf 70} (2004) 045203
  [arXiv:nucl-th/0402086].
\bibitem{mitandreas}
  A.~Nogga and C.~Hanhart,
  Phys.\ Lett.\ B {\bf 634} (2006) 210
  [arXiv:nucl-th/0511011].
\bibitem{Erics} 
 T.~E.~O.~Ericson, B.~Loiseau and A.~W.~Thomas,
  Phys.\ Rev.\  C {\bf 66} (2002) 014005
  [arXiv:hep-ph/0009312].
\bibitem{BBEMP}
S.~R.~Beane, V.~Bernard, E.~Epelbaum, U.-G.~Mei{\ss}ner and D.~R.~Phillips,
Nucl.\ Phys.\ A {\bf 720} (2003) 399
[arXiv:hep-ph/0206219].
\bibitem{BKM1}
 V.~Bernard, N.~Kaiser and U.-G.~Mei\ss ner,
  Nucl.\ Phys.\  A {\bf 615} (1997) 483
  [arXiv:hep-ph/9611253].
\bibitem{mrr_iso}
  U.-G.~Mei\ss ner, U.~Raha and A.~Rusetsky,
  Phys.\ Lett.\  B {\bf 639} (2006) 478
  [arXiv:nucl-th/0512035].
\bibitem{trento}
  C.~Curceanu, A.~Rusetsky and E.~Widmann,
  arXiv:hep-ph/0610201.
\bibitem{piploop}
 J.~Gasser, M.~A.~Ivanov, E.~Lipartia, M.~Moj\v zi\v s and A.~Rusetsky,
  Eur.\ Phys.\ J.\  C {\bf 26} (2002) 13
  [arXiv:hep-ph/0206068].
\bibitem{nadia2}
  N.~Fettes and U.-G.~Mei\ss ner,
  Nucl.\ Phys.\  A {\bf 693} (2001) 693
  [arXiv:hep-ph/0101030].



\end{thebibliography}
\end{document}